\documentstyle[12pt,aasms4]{article}
\slugcomment{PASP revised manuscript, 12 Aug 96, proofs to NZ} 
\lefthead{Zacharias, N.}
\righthead{Atmospheric Influence on Differential Astrometry}
\begin{document}

\title{Measuring the Atmospheric Influence on Differential Astrometry:\\
       a Simple Method Applied to Wide Field CCD Frames}
\author{N. Zacharias\altaffilmark{1}} 
\affil{U.S. Naval Observatory, 3450 Mass. Ave. N.W., Washington D.C. 20392,\\
       Electronic mail: nz@pyxis.usno.navy.mil}

\altaffiltext{1}{with Universities Space Research Association (USRA),
  Division of Astronomy and Space Physics, Washington D.C., 
         based on observations made at KPNO and CTIO}

\begin{abstract}
Sets of short exposure, guided CCD frames are used to measure 
the noise added by the atmosphere 
to differential astrometric observations.
Large nightly variations that are correlated with the seeing
have been found in the data obtained
over 2 years at the KPNO and CTIO 0.9--meter telescopes.
The rms noise added by the atmosphere,
after a linear transformation of the raw $x,y$ data, is found to be
3 to 7 mas, normalized to 100 seconds exposure time
and a field of view of 20 arcminutes near the zenith.
An additional nearly constant noise (base--level) of 
8.5 mas = 0.012 pixel is found for the KPNO and
6.0 mas = 0.015 pixel for the CTIO telescope. 
This implies that a ground--based, all sky, astrometric survey from 
guided CCD frames is more likely limited by the base--level noise 
than by the atmosphere and could reach an accuracy better than 10 mas
under good seeing conditions.

\end{abstract}


\section{INTRODUCTION}

Turbulence in the Earth's atmosphere adds noise
to ground--based astrometric observations. 
Semi--empirical and empirical results have been published previously, e.g.
(\cite{lin80}; \cite{kle83}; \cite{han89}; \cite{MM92}, \cite{hg95}). 
This effect ultimately limits the accuracy of ground--based astrometric
observations, and it is important to find these limits.
The effect is largest, about 100 mas, for absolute astrometry.
For differential astrometry, previous investigations have dealt with
{\em angular separation} measures.
The effect is found to be at the 1--2 mas level for arcminute separations
and several minutes integration time (\cite{hg95}),
e.g. applicable to double star and parallax observations. 

Here we will go one step further and define $\sigma_{atm}$ as the added noise
introduced by the atmosphere to astrometric observations, 
after an orthogonal or linear mapping model
has been applied to the $x,y$ data of guided exposures.
This is more appropriate for astrometric imaging observations,
because such a mapping model is used for the calibration of the $x,y$ data
to the reference star positions anyway, thus absorbing terms like
scale factor and field rotation.
The proposed technique in principle can be used with photographic plates as well 
as with CCD imaging, although the use of CCDs is more likely to show 
any atmospheric effect due to usually shorter exposure times and higher
internal precision.

The dependence of $\sigma_{atm}$ on integration time is well known to be
$\sigma_{atm} \ \sim \ t^{-1/2} $ 
and we assume this relationship here in our definition of $\sigma_{atm}$.
The dependence of $\sigma_{atm}$ on the field of view (FOV) is approximately known
to be $\sigma_{atm} \ \sim \ (FOV)^{-1/3}$
(e.g. \cite{han89}), at least for fields smaller than about half a degree,
and will not be investigated here.
Our goal is to determine the range of $\sigma_{atm}$ for 
different nights and atmospheric conditions and look for a dependence on seeing, 
as determined from the full width at half maximum (FWHM) of the image profiles.

\section{METHOD}

A simple method is introduced here to measure $\sigma_{atm}$ 
based on direct CCD imaging without the need for further instrumentation.
CCD frames have been taken of fields with a high star density and reduced
by standard procedures including bias removal and flat--fielding.
Circular symmetric 2--dimensional Gaussian image profiles have been 
fitted by least--squares methods to the flat--fielded CCD pixel data. 
Fig.~1 gives an example for the standard error in position plotted
vs.~instrumental magnitude for individual images.
% frame h10113, 2.night, 1901+319, 40 sec
Stars within a dynamic range from the saturation limit (here set to 10th magnitude)
to about 5 magnitudes fainter, display an almost constant level of precision 
for the $x,y$ position as obtained by the image profile fit.
The positional error increases for fainter stars because of the smaller S/N ratio
and for brighter stars because of the model insufficiencies
(saturated pixels, diffraction spikes).
Images well above the average position error for their magnitudes are either from double 
stars or galaxies and have been excluded from this investigation.
This diagram does not change with exposure time,
except for a shift along the magnitude scale and the number of images available
in a given range of instrumental magnitudes.

Assume 2 CCD frames of equal exposure times have been taken within a short period of 
time under the same conditions (atmosphere and telescope).
The field center of the second exposure has been shifted by a few pixels 
with respect to the first one.
Thus, independent observations have been obtained with images of the same star
located on different pixels of the CCD for both frames. 
Only the repeatability of the observations is investigated here,
so no attempt has been made to convert the $x,y$ measures into right ascensions
and declinations.
All error contributions related to field distortions are avoided because 
the same approximate field center has been used for both exposures.

The $x,y$ coordinates of the first frame are transformed into the system
of the $x,y$ coordinates of the second frame with a least--squares fit using either a
linear or orthogonal model.
Only those stars within the magnitude range of almost constant fit precision
as described above, have been used for this transformation.
The variance of the transformation between the 2 exposures, $\sigma_{trans}^{2}$, is

\[ \sigma_{trans}^{2} \ = \ 2 \ ( \sigma_{atm}^{2} \ + \ \sigma_{b}^{2} )  \]

with $\sigma_{atm}$ being the contribution from the atmosphere and
$\sigma_{b}$ the remaining error contribution -- the base--level -- as inherent
in our procedure and instrument (model insufficiencies, digitization errors, etc.),
for each individual CCD frame.
Defining $\sigma_{a}$ from $\sigma_{atm} \ = \ \sigma_{a} \ t^{-1/2} $
with exposure time $t$ in seconds we arrive at  

\[ \sigma_{trans}^{2} / 2 \ = \ \sigma_{a}^{2} \ t^{-1} \ + \sigma_{b}^{2}
			  \ = \ \sigma^{2}  \]

which is a linear relationship between the observable quantity $\sigma^{2}$
and the nearly error free parameter $t^{-1}$.
Assuming constant observational conditions for the time to take more sets
of CCD frame pairs for other exposure times, 
we can solve for $\sigma_{a}$ and $\sigma_{b}$.

\section{OBSERVATIONS}

Observing runs for the Radio--Optical Reference Frame (RORF) project (\cite{joh91})
have been conducted from 1994 to 1996 at the 0.9--meter telescopes on Kitt Peak
and Cerro Tololo (\cite{zac95}).
The KPNO 0.9--meter telescope has a scale of 0.68"/pixel and a FOV 
of 23.2', while those numbers are 0.40"/pixel and 13.6' for the
CTIO telescope. 

A few CCD frames were specifically taken 
to investigate the limits set by the atmosphere on astrometric accuracy.
Fields with a high star density (close to the Milky Way) and close to the 
zenith, if possible, were observed close to the meridian.
For most of the selected fields, 2 frames of 40, 20 and 10 seconds
exposure time each were taken with the same Gunn r filter in addition to
the long exposures for the RORF project.
An off--axis autoguider was used with guide stars selected close to the
edge of the main FOV.
The integration and correction cycle time was set to about 1 second
and the system was guiding for at least 10 seconds before the start of
a new exposure.

Figure 2 shows an example of $\sigma^{2}$  plotted vs. $t^{-1}$
for the 4 exposure times. 
A linear model has been used for the $x,y$ transformations.
The filled circles and open squares are the results for the $x$ and $y$ coordinates
respectively (along $\delta$, $\alpha$ for the KPNO telescope). 

Because there are more faint than bright stars,
the longer exposure frames show more stars
near the saturation limit than the short exposure frames.
Also, for a given constant number of stars used for different transformations,
the value of $\sigma_{trans}$ is better determined for longer exposure times
because of the smaller scatter in the $x,y$ transformation data
due to better image definition from the longer integration time.
Thus, weights have been assigned to each measured $\sigma_{trans}^{2}$ value.
Let $E(y)$ be an estimate of the standard error of the quantity $y$ and
$y = x^{2}$ with $x = \sigma_{trans}$, then we have from the error
propagation law

\[  E(y) \ = \ E(x^{2}) \ = \ 2 \ x \ E(x)  \]

$E(x)$ is here the standard error of the mean, using all $n_{stars}$
star pairs for the transformation, thus

\[  E(x) \ = \ \sigma_{trans} / \sqrt{n_{stars}}       \]

Putting everything together we arrive at the adopted formula

\[  error \ estimate \ on \ \sigma_{trans}^{2} \ = \ 2 \ \sigma_{trans}
     \ \frac{\sigma_{trans}}{\sqrt{n_{stars}}}       \]

The weighting is not critical for the determination of the slope itself, i.e.
for the atmospheric contribution.
The determination of the base--level and the error estimates on the results
depend more strongly on the choice of the weighting algorithm. 
This conclusion was obtained from tests made with different weighting
algorithms, including unweighted reductions.

A weighted least--squares fit was performed with the 
$\sigma^{2}$ vs. $t^{-1}$ data points,
in order to obtain the slope and constant of the straight line predicted
by the theory. Fit results for each axis separately (dotted, full line)
as well as for the combined data (dashed line), are shown in Fig.~2.

\section{RESULTS}

From the straight line fit of the $\sigma^{2}$ vs. $t^{-1}$ plots
for the combined data of both axes,
$\sigma_{a}$ and $\sigma_{b}$, and their errors were calculated.
Table 1 lists all observations and results. 
The first line for each observation shows the result from the linear
transformation model, while the second line shows the result as obtained
with the orthogonal model. 
The last column displays $\sigma_{an}$, which is $\sigma_{a}$ normalized 
to 100 seconds exposure time and a FOV of 20 arcminutes for the zenith,
obtained from

\[  \sigma_{an} \ = \ \sigma_{a} \ \left( \frac{1^{s}}{100^{s}} \right)^{1/2} \ 
		      \left( \frac{20'}{fov} \right)^{1/3} \
		      \cos z                    \] 

with $fov$ being the field of view in arcminutes as used for the
$x,y$ transformations and $z$ being the mean zenith distance while
observing the field.
Fig.~3 shows results for $\sigma_{an}$ obtained with the linear
$x,y$ transformation model plotted vs. FWHM, scaled to the zenith with
a $\cos z$ factor, adopted from (\cite{lin80}).

\section{DISCUSSION} 

There is a large nightly variation in the atmospheric influence on
the observed star positions, which is correlated
with seeing (FWHM), but "the seeing value" alone 
does not determine $\sigma_{an}$.
On the average we obtain $\sigma_{an} \ \approx$ 3 mas and 6 mas for 1 
and 2 arcsecond seeing respectively.

The results obtained with the orthogonal $x,y$ transformation model
give on the average larger numbers for $\sigma_{an}$ by about 10\%.
This is expected, because fewer parameters will model less of the 
real atmospheric effects.

All our results hold only for this type of guided imaging
observing procedure. For differential transit circle 
or scanning mode observations, the modelling of the atmospheric 
effects is different, as will be the residual effects caused by
the atmosphere on the astrometric results (\cite{sto96}).

Lindegren (1980) obtained standard errors for observing the separation
of stars, i.e. a different kind of differential astrometry from that 
investigated here. His result, scaled to 100 seconds exposure time
and a mean separation of 10' (comparable to our 20' FOV),
is about 19 mas.
Results by Han (1989) would lead to 14 mas for this case.
Both Lindegren's and Han's results are obtained in medium
seeing conditions ($\approx 2"$), thus they have to be compared to
our 6 mas, which is a factor of 2 to 3 smaller.
Scaled to a 100 second exposure time and a star separation of 10',
Han and Gatewood (1995) found $\sigma_{an}$ = 5.4 mas from star trail 
observations obtained from Mauna Kea.
Our result is 3 mas for good seeing, which is smaller by
almost a factor of 2. 

Separation measurements made at the 61--inch Flagstaff telescope
result in an atmospheric contribution of 9 to 28 mas for this case,
depending on the night (\cite{MM92}, \cite{M96}).
Again our result is a factor of 2 to 3 smaller. 
Similar to our results, Monet and Monet find a lose correlation with seeing,
which ranged from FWHM=1.2" to 2.3". 
According to a hypothesis (\cite{M96}), local effects near the dome
cause some of the "astrometric seeing", not correlated with the general FWHM.

This difference between our results and those obtained by other investigations
can be explained by the different types of observations.
The {\em simultaneous} observation of all stars in the FOV seems to be
important.
Moreover, our guided exposures follow correlated image motions of a star field,
caused by the atmosphere, and thus reduce the noise contribution 
compared to other differential observing procedures.
Also, a linear reduction model will give smaller residuals as 
compared to angular separation measurements with fewer free parameters.

As a by--product of this method, the base--level of accuracy has
been determined as well.
The mean of all $\sigma_{b}$ with a standard error smaller than 1.0 mas
is found to be  8.5 mas = 0.012 pixel for all KPNO observations. 
The corresponding result for the CTIO instrument is 6.0 mas = 0.015 pixel.
These numbers are for a single observation per coordinate.
The slightly smaller value for $\sigma_{b}$ (in pixel units) for the
KPNO instrument can be explained by the better optical quality of that
telescope, which includes a field flattener corrector lens. 
Imperfections in the CCD, the optics of the telescope and model
deficiencies contribute to this base--level of astrometric accuracy,
which is under further investigation 
(\cite{zr95}; \cite{win96}; \cite{zac97}).

\section{CONCLUSIONS}

A large nightly variation (factor of 2) of the noise added by the 
atmosphere to differential astrometry has been found.  
This added noise is correlated with the seeing.
In good seeing conditions ($\approx 1"$) the contribution of the 
atmosphere to differential astrometry can be as small as 3 mas for
guided 100--second exposures and a FOV of 20 arcminutes for 0.9--meter
aperture telescopes.

Guided exposures with simultaneous observation of all stars in the FOV
give a considerable (about a factor of 2) advantage over
angular separation measurements made with other differential astrometric
observing techniques.

For a 1 degree FOV astrometric survey telescope, our result scales to 
4.3 mas noise contributed by the atmosphere for 100--second exposures.
This is considerably less than previously expected.
Thus the limit to ground--based wide field astrometric observations 
as set by the atmosphere has not yet been reached for long integration
times ($\ge 100$ seconds) because of the relatively 
large base--level noise of $\approx 0.015 $ pixels, which is on the order 
of 6 to 15 mas, depending on the sampling.

\acknowledgments

I would like to thank Kitt Peak and Cerro Tololo Observatories
for granting observing time.
I am grateful to M.I.Zacharias for assistance in observing
and reduction of the CCD frames.

\newpage

\begin{figure}
\caption{Precision in the y--coordinate (along $\alpha$) for star image profile fits of
   a typical CCD frame vs. instrumental magnitude.
   This example is from a 40 second exposure obtained at the KPNO 0.9--meter telescope
   in 1.6" seeing. The scale is 0.01 pixel = 6.8 mas.}
\end{figure}

\begin{figure}
\caption{Variance of $x,y$ transformation ($\sigma^{2} \ = \ \sigma_{trans}^{2}/2$)
   in $mas^{2}$ vs. inverse exposure time in $s^{-1}$ for an example from KPNO observations.
   The filled circles are for the x--coordinate ($\delta$) and the open boxes are for
   the y--coordinate ($\alpha$). Fit results are shown as dotted, full and dashed lines
   (y only, x only, both coordinates).}
\end{figure}

\begin{figure}
\caption{Error contribution by the atmosphere to differential astrometry 
   vs. FWHM of image profiles, scaled to the zenith. Full circles and open boxes 
   show results from the KPNO and CTIO telescopes respectively.
   Only results of the linear $x,y$ transformation model are shown here.} 
\end{figure}

\newpage

\begin{table}
\begin{center}
TABLE 1 \\
Observations and results \\

\vspace*{5mm}

\begin{tabular}{lrrrrrrrrr}
\tableline
\tableline
tel. &  date  &  z      &  FWHM   & nexp & $\sigma_{a}$& error & $\sigma_{b}$ & error & $\sigma_{an}$ \\
	  & (ymd)  & (degree)& (arcsec)&      & (mas)       & (mas) & (mas) & (mas)  & (mas) \\
\tableline
%start data    this line is req. for pgm fig3
K      & 940420 &   6     &  2.00   &   4  &  62   &   9  &  7.0  &  0.5   &  5.9  \\
       &        &         &         &      &  74   &   6  &  7.5  &  0.3   &  7.1  \\
K      & 940422 &  21     &  2.10   &   4  &  39   &  15  & 10.2  &  1.0   &  3.6  \\
       &        &         &         &      &  50   &  16  & 10.3  &  1.2   &  4.5  \\
K      & 940423 &  17     &  2.65   &   3  &  68   &  13  &  8.2  &  3.0   &  6.3  \\
       &        &         &         &      &  60   &  15  & 10.1  &  2.9   &  5.5  \\

K      & 941020 &  44     &  1.70   &   4  &  43   &  12  &  9.1  &  0.9   &  3.1  \\
       &        &         &         &      &  35   &  20  & 12.0  &  1.2   &  2.5  \\

K      & 950613 &   1     &  1.65   &   4  &  31   &   7  &  8.8  &  0.5   &  3.0  \\
       &        &         &         &      &  36   &  10  &  9.0  &  0.7   &  3.5  \\
K      & 950615 &  12     &  2.10   &   4  &  72   &   6  &  9.1  &  0.5   &  6.8  \\
       &        &         &         &      &  80   &   7  &  8.9  &  0.6   &  7.5  \\
K      & 950617 &  15     &  1.30   &   4  &  38   &   6  &  7.9  &  0.5   &  3.6  \\
       &        &         &         &      &  39   &   7  &  8.0  &  0.6   &  3.7  \\

K      & 960105 &  15     &  1.20   &   5  &  39   &   8  &  8.8  &  0.9   &  3.6  \\
       &        &         &         &      &  45   &   7  &  9.2  &  0.9   &  4.1  \\

C      & 941215 &  17     &  1.35   &   5  &  54   &   7  &  5.8  &  0.7   &  6.0  \\
       &        &         &         &      &  65   &  10  &  7.4  &  1.1   &  7.3  \\

C      & 950213 &   1     &  1.30   &   4  &  39   &   6  &  6.5  &  0.6   &  4.4  \\
       &        &         &         &      &  75   &   4  &  9.2  &  0.5   &  8.5  \\

C      & 950917 &   1     &  1.60   &   4  &  67   &   6  &  5.9  &  0.8   &  7.6  \\
       &        &         &         &      &  72   &  15  &  7.8  &  1.6   &  8.2  \\
\tableline
\tableline
\end{tabular}
\end{center}

\end{table}

\end{document}